\documentclass[aps,prb,twocolumn,superscriptaddress]{revtex4-1}

\usepackage{graphicx}
\usepackage{dcolumn}
\usepackage{bm}
\usepackage[version=3]{mhchem}

\begin{document}

\title{The role of native defects in the transport of charge and mass and the decomposition of Li$_4$BN$_3$H$_{10}$}
\author{Khang Hoang}
\affiliation{Center for Computationally Assisted Science and Technology, North Dakota State University, Fargo, ND 58108, USA.}
\author{Anderson Janotti}
\affiliation{Materials Department, University of California, Santa Barbara, California 93106, USA.}
\author{Chris G. Van de Walle}
\affiliation{Materials Department, University of California, Santa Barbara, California 93106, USA.}


\begin{abstract}

Li$_{4}$BN$_{3}$H$_{10}$ is of great interest for hydrogen storage and for lithium-ion battery solid electrolytes because of its high hydrogen content and high lithium-ion conductivity, respectively. The practical hydrogen storage application of this complex hydride is, however, limited due to irreversibility and cogeneration of ammonia (NH$_{3}$) during the decomposition. We report a first-principles density-functional theory study of native point defects and defect complexes in Li$_{4}$BN$_{3}$H$_{10}$, and propose an atomistic mechanism for the material's decomposition that involves mass transport mediated by native defects. In light of this specific mechanism, we argue that the release of NH$_{3}$ is associated with the formation and migration of negatively charged hydrogen vacancies inside the material, and it can be manipulated by the incorporation of suitable electrically active impurities. We also find that Li$_{4}$BN$_{3}$H$_{10}$ is prone to Frenkel disorder on the Li sublattice; lithium vacancies and interstitials are highly mobile and play an important role in mass transport and ionic conduction.

\end{abstract}


\maketitle


\section{Introduction}

The ability to store hydrogen for subsequent use is key to a hydrogen economy where hydrogen serves as an energy carrier in a carbon-neutral energy system.\cite{Eberle2009} Complex hydrides such as Li$_{4}$BN$_{3}$H$_{10}$ have been considered for hydrogen storage because of their high theoretical hydrogen density.\cite{orimo_chem_rev_2007} Li$_{4}$BN$_{3}$H$_{10}$, which is synthesized from mixtures of LiBH$_{4}$ and LiNH$_{2}$ in a 3:1 molar ratio, releases greater than 10 wt\% hydrogen when heated.\cite{meisner2006} Yet its practical application is limited due to the cogeneration of ammonia (NH$_{3}$) and the irreversibility of the decomposition reaction.\cite{Pinkerton2006,Liu2013} It has also been reported that metal additives such as NiCl$_{2}$, Pd (or PdCl$_{2}$), and Pt (or PtCl$_{2}$) can suppress the release of NH$_{3}$ gas from Li$_{4}$BN$_{3}$H$_{10}$ and lower the dehydrogenation temperature;\cite{Pinkerton2006,Pinkerton2007} however, the role of these additives is still not well understood. In addition to hydrogen storage, Li$_{4}$BN$_{3}$H$_{10}$ has also shown promise as a battery solid electrolyte due to its high lithium-ion conductivity.\cite{MatsuoReview} The material was reported to have a conductivity of $2\times 10^{-4}$ S/cm at room temperature and an activation energy of 0.26 eV.\cite{Matsuo2009}

The decomposition of Li$_{4}$BN$_{3}$H$_{10}$ can proceed as\cite{meisner2006,Herbst2006}
\begin{equation}\label{eq:pathway}
\mathrm{Li}_{4}\mathrm{B}\mathrm{N}_{3}\mathrm{H}_{10} \rightarrow \mathrm{Li}_{3}\mathrm{B}\mathrm{N}_{2} + \frac{1}{2}\mathrm{Li}_{2}\mathrm{NH} + \frac{1}{2}\mathrm{NH}_{3} + 4\mathrm{H}_{2}.
\end{equation}
While other reaction pathways have been proposed,\cite{Herbst2006,Siegel2007} reaction (\ref{eq:pathway}) whose products contain 8.9 and 9.4 mass \% of H$_{2}$ and NH$_{3}$ is considered to be closest to the experimental situation where the values of 9.6 and 8.4 mass \%, respectively, have been observed.\cite{meisner2006,Herbst2006} The decomposition and dehydrogenation of Li$_{4}$BN$_{3}$H$_{10}$, like those of other complex hydrides such as LiBH$_{4}$ and LiNH$_{2}$, necessarily involves the breaking and forming of chemical bonds and the transport of mass in the bulk. These are electronic and atomistic processes that can be fruitfully studied using first-principles calculations.\cite{Peles2007,wilson-short,Hao2010,Miceli2010,Wang2011,hoang2011LiBH4} Comprehensive and systematic computational studies of the structure, energetics, and migration of native point defects and defect complexes can provide direct insights into the mechanisms for decomposition and dehydrogenation, help in identifying the rate-limiting processes, and ultimately aid in design of materials with improved hydrogen desorption kinetics.

Native point defects in Li$_{4}$BN$_{3}$H$_{10}$ were first studied by us based on density-functional theory (DFT);\cite{hoang_prb_2009} however only hydrogen-related defects were considered. A more comprehensive study was carried out by Farrell and Wolverton;\cite{farrell2012} they reported not only results for hydrogen vacancies and interstitials but also for some lithium-, boron-, and nitrogen-related defects. They also studied the dependence of defect formation energies on the atomic chemical potentials. Farrell and Wolverton,\cite{farrell2012} however, considered neither defect complexes nor migration of the native defects that may play an important role in mass and charge transport in the material. As reported in previous work,\cite{wilson-short,hoang2011LiBH4,hoang2011amideAC} Frenkel defect pairs, i.e., interstitial-vacancy complexes of the same species, can play an essential role in decomposition and dehydrogenation processes.

Here we report a comprehensive and systematic DFT study of the structure, energetics, and migration of hydrogen-, lithium-, boron-, and nitrogen-related isolated native point defects in all the possible charge states, as well as defect complexes in Li$_{4}$BN$_{3}$H$_{10}$. Some results for hydrogen-related defects were reported previously,\cite{hoang_prb_2009} but are included here after applying finite-size effect corrections to the formation energy of charged hydrogen vacancies and interstitials (see details in Sec.~2); a lower energy configuration of the neutral hydrogen interstitial is also reported. We find that Li$_{4}$BN$_{3}$H$_{10}$ is prone to Frenkel disorder on the Li sublattice, and the lithium vacancies and interstitials are highly mobile and can play an important role in mass transport and ionic conduction. On the basis of our results, we propose a specific mechanism for the decomposition of Li$_{4}$BN$_{3}$H$_{10}$ in which the release of NH$_{3}$ is associated with the formation and migration of negatively charged hydrogen vacancies in the interior of the material. In light of this mechanism, we discuss the role of transition metal impurities such as Ni, Pd, and Pt in suppressing the release of NH$_{3}$ and in lowering the dehydrogenation temperature. Comparison with previous computational studies will be made where appropriate.

\section{Methodology}

Our calculations are based on DFT using the generalized-gradient approximation~\cite{GGA} and the projector-augmented wave method,~\cite{PAW1,PAW2} as implemented in the Vienna {\it Ab Initio} Simulation Package (VASP).~\cite{VASP1,VASP2,VASP3} We used the unit cell of Li$_{4}$BN$_{3}$H$_{10}$ containing 144 atoms and a 2$\times$2$\times$2 Monkhorst-Pack $\mathbf{k}$-point mesh.~\cite{monkhorst-pack} The plane-wave basis-set cutoff was set to 400 eV. Convergence with respect to self-consistent iterations was assumed when the total energy difference between cycles was less than 10$^{-4}$ eV and the residual forces were less than 0.01 eV/{\AA}. In the defect calculations, the lattice parameters were fixed to the calculated bulk values, but all the internal coordinates were fully relaxed. Migration was studied using the climbing-image nudged elastic band (NEB) method.~\cite{ci-neb}

We characterize different defects in Li$_{4}$BN$_{3}$H$_{10}$ using their formation energies. Defects with low formation energies will easily form and occur in high concentrations. The formation energy of a defect X in charge state $q$ is defined as~\cite{walle:3851}
\begin{equation}\label{eq:eform}
E^f({\mathrm{X}}^q)=E_{\mathrm{tot}}({\mathrm{X}}^q)-E_{\mathrm{tot}}({\mathrm{bulk}})-\sum_{i}{n_i\mu_i}+q(E_{\mathrm{v}}+\mu_{e})+ \Delta^q ,
\end{equation}
where $E_{\mathrm{tot}}(\mathrm{X}^{q})$ and $E_{\mathrm{tot}}(\mathrm{bulk})$ are, respectively, the total energies of a supercell containing the defect X and of a supercell of the perfect bulk material; $\mu_{i}$ is the atomic chemical potential of species $i$ (and is referenced to bulk Li metal, bulk B metal, N$_{2}$ molecules, or H$_{2}$ molecules at 0 K), and $n_{i}$ denotes the number of atoms of species $i$ that have been added ($n_{i}>$0) or removed ($n_{i}<$0) to form the defect. $\mu_{e}$ is the electronic chemical potential, i.e., the Fermi energy, referenced to the valence-band maximum in the bulk ($E_{\mathrm{v}}$). $\Delta^q$ is the correction term to align the electrostatic potentials of the bulk and defect supercells and to account for finite-cell-size effects on the total energies of charged defects.\cite{walle:3851} To correct for the finite-size effects, we adopted the Freysoldt {\it et al.}'s approach,\cite{Freysoldt,Freysoldt11} using a static dielectric constant of 15.38 calculated using density functional perturbation theory.~\cite{Wu2005,dielectricmethod} In our calculations, $\Delta^q$ can be as low as 0.05 eV for a singly charged defect or as high as 1.40 eV for a triply charged defect. We note that in Ref.~\cite{hoang_prb_2009} $\Delta^q = 0$, i.e., no corrections were included in the results reported there. Farrell and Wolverton, on the other hand, took into account only the ``potential alignment'' term, reported to be of $\sim$0.2$-$0.9 eV depending specific defects.\cite{farrell2012}

The chemical potentials $\mu_{i}$ are variables and can be chosen to represent experimental situations. For defect calculations in Li$_{4}$BN$_{3}$H$_{10}$, from Eq.~(\ref{eq:pathway}) an equilibrium between Li$_{3}$BN$_{2}$, Li$_{2}$NH, and Li$_{4}$BN$_{3}$H$_{10}$ can be assumed and the chemical potentials of Li, B, and N are expressed in terms of $\mu_{\mathrm{H}}$, which is now the only variable. In the following presentation, we set $\mu_{\mathrm{H}}=-0.15$ eV, corresponding to the Gibbs free energy of H$_{2}$ gas at 1 bar and 282 K.\cite{H2gas} This condition gives $\mu_{\mathrm{Li}}=-0.24$ eV, $\mu_{\mathrm{B}}=-1.51$ eV, and $\mu_{\mathrm{N}}=-1.47$ eV. With this set of chemical potentials, the calculated formation energies of the defects in Li$_{4}$BN$_{3}$H$_{10}$ are all non-negative, as shown in the next section. One can choose a different set of chemical potentials and that may affect the formation energies; however, our conclusions should not depend on the chemical potential choice. We also note that the Fermi energy $\mu_{e}$ is not a free parameter but subject to the charge-neutrality condition that involves all possible native defects and any impurities present in the material.\cite{walle:3851}

\section{Results}

Li$_{4}$BN$_{3}$H$_{10}$ was reported to crystallize in the cubic space group $I$2$_{1}$3.~\cite{filinchuk,yang} This quaternary compound can be considered as a mixture of end compounds LiBH$_{4}$ and LiNH$_{2}$, or an ionic compound in which (Li)$^{+}$, (NH$_{2}$)$^{-}$, and (BH$_{4}$)$^{-}$ units are the basic building blocks. The valence-band maximum (VBM) of Li$_{4}$BN$_{3}$H$_{10}$ consists of nitrogen-related unbonded states coming from the (NH$_{2}$)$^{-}$ units, whereas the conduction-band minimum (CBM) is composed of a mixture of N $p$ and H $s$ states. The electronic structure near the band-gap region is, therefore, dominated by that of LiNH$_{2}$.~\cite{hoang2011amideAC,hoang2011amidePRB} The calculated band gap is 3.53 eV$-$a direct gap at the $\Gamma$ point.\cite{hoang_prb_2009} Given the structural and electronic properties of Li$_{4}$BN$_{3}$H$_{10}$, one expects that native point defects in this compound will possess characteristics of those in the end compounds LiBH$_{4}$ and LiNH$_{2}$.~\cite{hoang2011amideAC,hoang2011amidePRB,hoang2011LiBH4}

\subsection{Hydrogen-related defects}

\begin{figure}
\begin{center}
\includegraphics[width=3.0in]{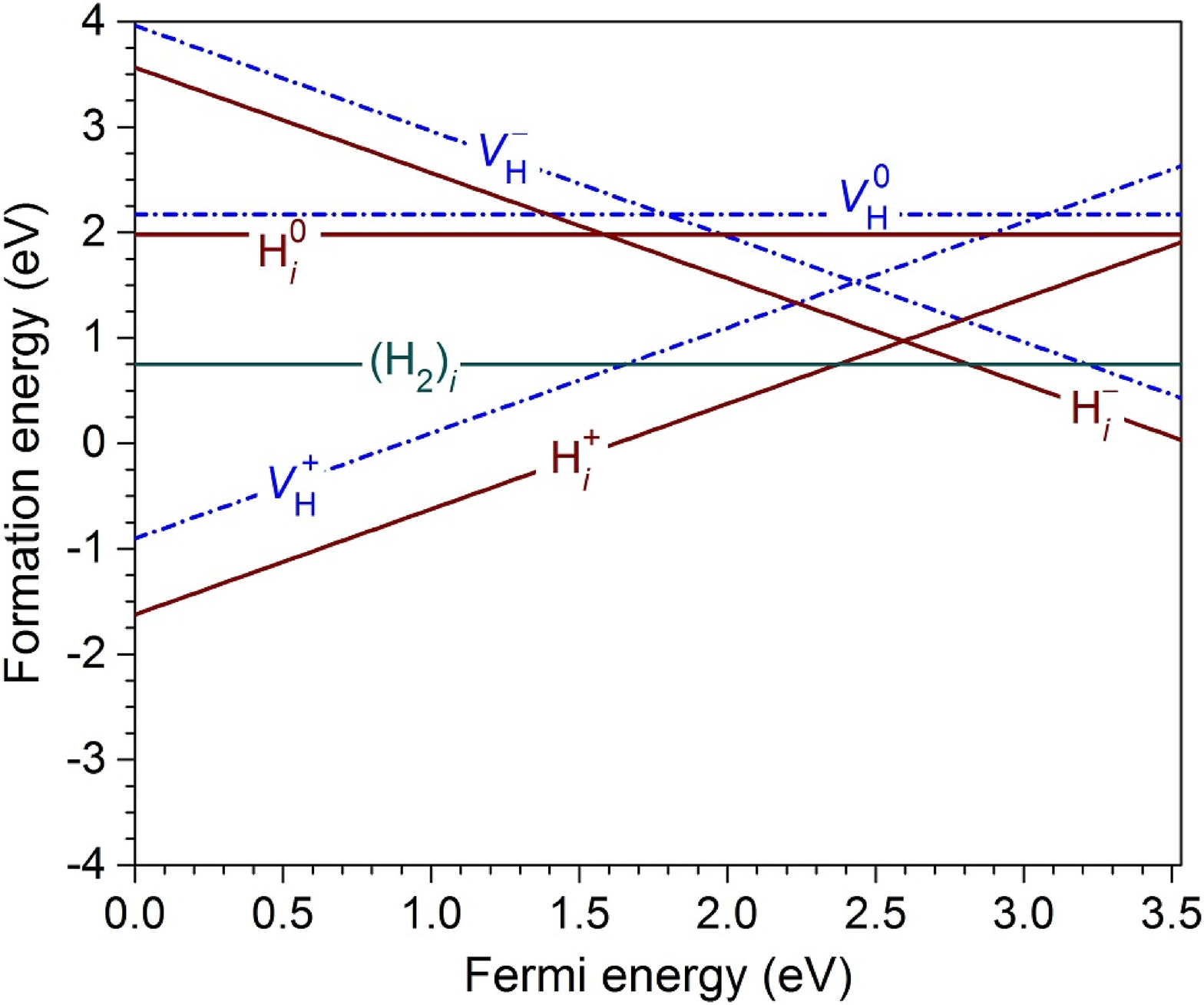}
\end{center}
\vspace{-0.2in}
\caption{Calculated formation energies of hydrogen-related defects, plotted as a function of Fermi energy with respect to the VBM.}\label{fig1}
\end{figure}

\begin{figure*}
\begin{center}
\includegraphics[width=6.5in]{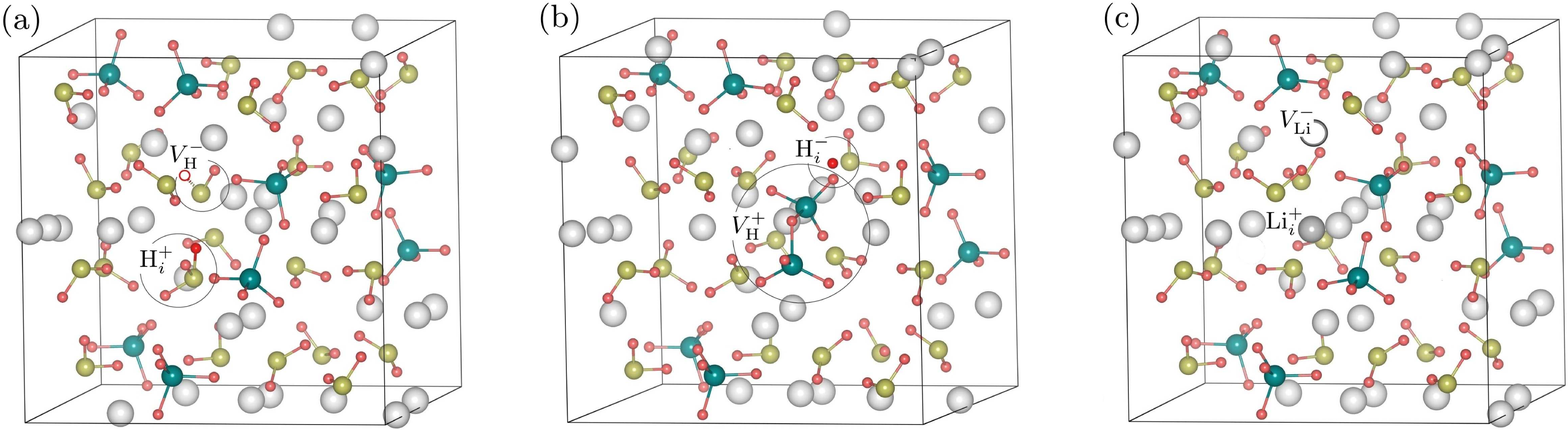}
\end{center}
\vspace{-0.15in}
\caption{Frenkel defect pairs: (a) (H$_{i}^{+}$,$V_{\mathrm{H}}^{-}$), (b) (H$_{i}^{-}$,$V_{\mathrm{H}}^{+}$), and (c) (Li$_{i}^{+}$,$V_{\mathrm{Li}}^{-}$). Large (gray) spheres are Li, medium (blue) spheres B, small (yellow) spheres N, and smaller (red) spheres H. Vacancies are represented by empty spheres.}\label{fig2}
\end{figure*}

Figure \ref{fig1} shows the calculated formation energies of hydrogen vacancies ($V_{\mathrm{H}}$), hydrogen interstitials (H$_{i}$), and a hydrogen molecule interstitial (H$_{2}$)$_{i}$. Among these defects, H$_{i}^{+}$ and H$_{i}^{-}$ have the lowest formation energies over a wide range of Fermi-energy values, except near their transition level, $\mu_{e}=2.59$ eV, where (H$_{2}$)$_{i}$ has the lowest formation energy.

We find that $V_{\mathrm{H}}^{0}$ is energetically most favorable when created by removing an H atom from a (BH$_{4}$)$^{-}$ unit, resulting in a trigonal planar BH$_{3}$. Similarly, $V_{\mathrm{H}}^{+}$ is created by removing a H atom and an extra electron from a (BH$_{4}$)$^{-}$ unit, forming a BH$_{3}$$-$H$-$BH$_{3}$ complex, i.e., two BH$_{4}$ units sharing a common H atom. $V_{\mathrm{H}}^{-}$, on the other hand, is most energetically favorable when created by removing an H$^{+}$ ion from a (NH$_{2}$)$^{-}$ unit, leaving the system with a (NH)$^{2-}$ unit.

For the interstitials, H$_{i}^{+}$ is most favorable when the added H$^{+}$ ion combines with an (NH$_{2}$)$^{-}$ unit to form an NH$_{3}$ unit. H$_{i}^{-}$, on the other hand, is found situated in a void surrounded by (Li)$^{+}$ units. Like H$_{i}^{+}$, H$_{i}^{0}$ also forms an NH$_{3}$ unit, which is in agreement with the configuration reported by Farrell and Wolverton,\cite{farrell2012} and is lower in energy than that reported previously by us where the neutral H atom loosely bonds to a (BH$_{4}$)$^{-}$ unit.\cite{hoang_prb_2009} However, even with this low-energy configuration, H$_{i}^{0}$ is never the most stable charge state of hydrogen interstitials. Finally, (H$_{2}$)$_{i}$, created by adding an H$_{2}$ molecule to the system, prefers to stay in the void formed by other species with the calculated H$-$H bond length being 0.76 {\AA}, which is comparable to that of an isolated H$_{2}$ molecule (0.75 {\AA}).

Our results thus indicate that in Li$_{4}$BN$_{3}$H$_{10}$, $V_{\mathrm{H}}^{0}$ and $V_{\mathrm{H}}^{+}$  possess the characteristics of those in LiBH$_{4}$; H$_{i}^{0}$, H$_{i}^{+}$, and $V_{\mathrm{H}}^{-}$ are similar to those in LiNH$_{2}$; and H$_{i}^{-}$ and (H$_{2}$)$_{i}$ inherit their structures from those in both the end compounds.\cite{hoang2011amideAC,hoang2011amidePRB,hoang2011LiBH4}

For the diffusion of H$_{i}^{+}$, H$_{i}^{-}$, $V_{\rm{H}}^{+}$, and $V_{\rm{H}}^{-}$, we find energy barriers of 0.48, 0.49, 0.64, and 1.02 eV, respectively. The barriers for H$_{i}^{+}$, $V_{\rm{H}}^{+}$, and $V_{\rm{H}}^{-}$ are higher because their diffusion involves breaking B$-$H or N$-$H bonds. For example, the diffusion of $V_{\rm{H}}^{-}$ involves moving a hydrogen atom from a nearby NH$_{2}$ unit to the vacancy. The saddle-point configuration in this case consists of a hydrogen atom located midway between two NH units, i.e., NH$-$H$-$NH. H$_{i}^{-}$, on the other hand, loosely bonds to (Li)$^{+}$ units and therefore can diffuse more easily. For comparison, the migration barriers of H$_{i}^{+}$, H$_{i}^{-}$, and $V_{\rm{H}}^{-}$ in LiNH$_{2}$ are 0.61, 0.34, and 0.71 eV,\cite{hoang2011amidePRB} and those of H$_{i}^{-}$ and $V_{\rm{H}}^{+}$ in LiBH$_{4}$ are 0.41 and 0.91 eV, respectively.~\cite{hoang2011LiBH4}

Possible hydrogen-related Frenkel defect pairs are (H$_{i}^{+}$,$V_{\mathrm{H}}^{-}$) and (H$_{i}^{-}$,$V_{\mathrm{H}}^{+}$). Figure~\ref{fig2}(a) shows the structure of (H$_{i}^{+}$,$V_{\mathrm{H}}^{-}$). The configurations of the individual defects are preserved in this complex, i.e., a NH$_{3}$ unit for H$_{i}^{+}$ and a (NH)$^{2-}$ unit for $V_{\mathrm{H}}^{-}$. The distance between the two N ions in the pair is 3.06 {\AA}. This Frenkel pair has a formation energy of 1.66 eV (independent of the chemical potentials), and a binding energy of 0.68 eV with respect to its isolated constituents. For comparison, a similar hydrogen Frenkel pair in LiNH$_{2}$ has a calculated formation energy of 1.54 eV.\cite{hoang2011amidePRB} Figure~\ref{fig2}(b) shows the structure of (H$_{i}^{-}$,$V_{\mathrm{H}}^{+}$). The configurations of the individual defects are also preserved in this case. The distance from H$_{i}^{-}$ to the H atom near the center of $V_{\mathrm{H}}^{+}$ is 4.25 {\AA}. This pair has a formation energy of 2.14 eV and a binding energy of 0.53 eV. The formation energy of this Frenkel pair in LiBH$_{4}$ is 2.28 eV.\cite{hoang2011LiBH4}

\subsection{Lithium-related defects}

\begin{figure}
\begin{center}
\includegraphics[width=3.0in]{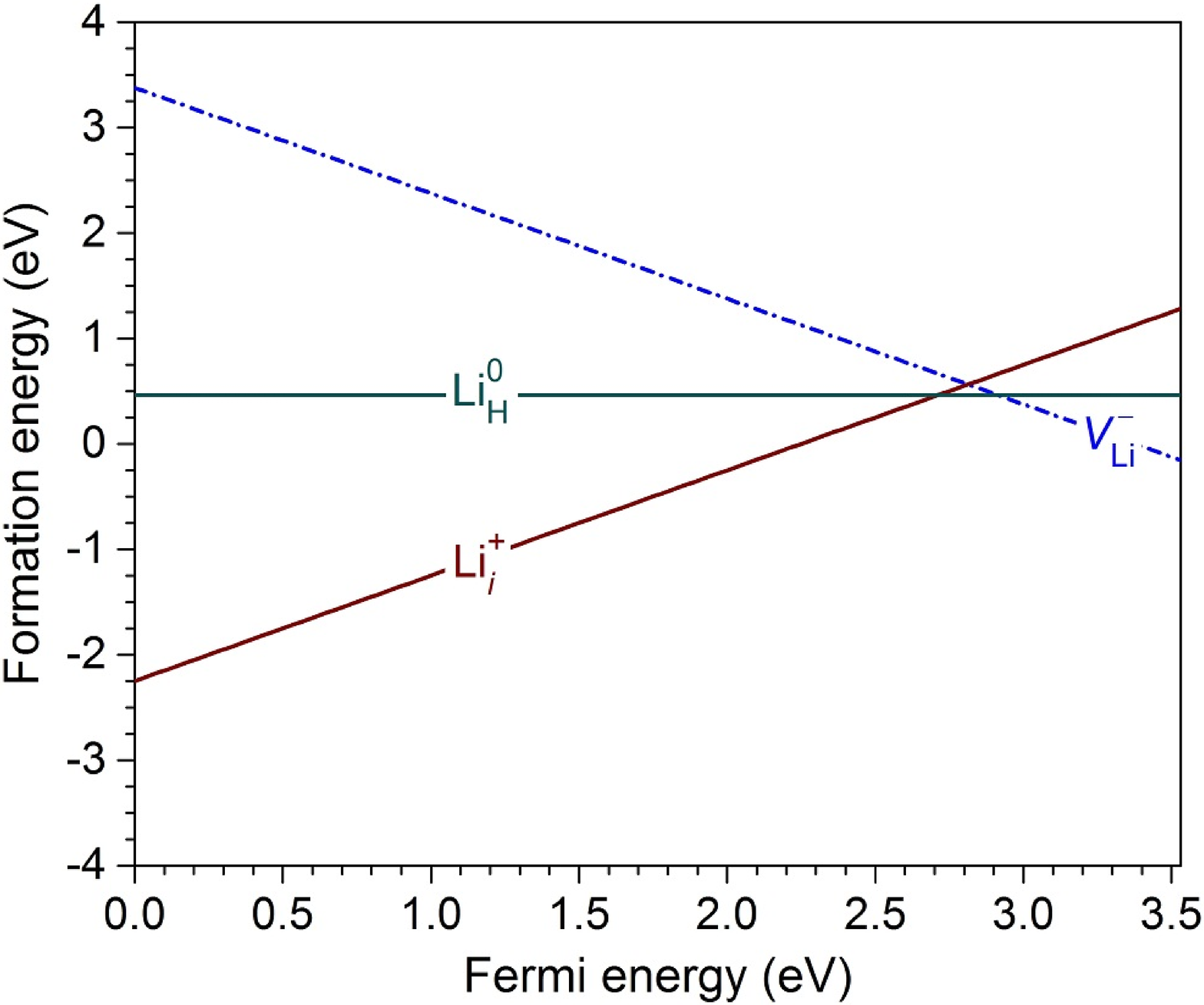}
\end{center}
\vspace{-0.2in}
\caption{Calculated formation energies of lithium-related defects, plotted as a function of Fermi energy with respect to the VBM.}\label{fig3} 
\end{figure}

Figure \ref{fig3} shows the calculated formation energies of lithium vacancies ($V_{\mathrm{Li}}$), interstitials (Li$_{i}$), and antisite defects (Li$_{\mathrm{H}}^{0}$, i.e., Li replacing an H atom). The creation of $V_{\mathrm{Li}}^{-}$ corresponds to the removal of a (Li)$^{+}$ unit from the system; whereas Li$_{i}^{+}$ can be thought of as the addition of a Li$^{+}$ ion to the system. These two defects result in relatively small local perturbations in the Li$_{4}$BN$_{3}$H$_{10}$ lattice. The creation of Li$_{\mathrm{H}}^{0}$, on the other hand, leaves the system with an (NH)$^{2-}$ unit and a Li interstitial. Thus, Li$_{\mathrm{H}}^{0}$ can be regarded as a complex of Li$_{i}^{+}$ and $V_{\mathrm{H}}^{-}$ with a binding energy of 1.25 eV. Since the resulting defects are a (NH)$^{2-}$ unit and a Li interstitial, the region that includes Li$_{\mathrm{H}}^{0}$ can be considered as locally Li$_{2}$NH inside the bulk Li$_{4}$BN$_{3}$H$_{10}$. This situation is similar to Li$_{\mathrm{H}}^{0}$ in LiNH$_{2}$.\cite{hoang2011amidePRB}

We find that Li$_{i}^{+}$ and $V_{\mathrm{Li}}^{-}$ have migration barriers of 0.43 and 0.20 eV, respectively. For Li$_{\mathrm{H}}^{0}$, which is considered as a complex of Li$_{i}^{+}$ and $V_{\mathrm{H}}^{-}$, the lower bound of the barrier is 1.02 eV, i.e., given by the least mobile species.\cite{hoang2011amidePRB} For comparison, the migration barrier of Li$_{i}^{+}$ in LiNH$_{2}$ and LiBH$_{4}$ is 0.30 eV, and that of $V_{\mathrm{Li}}^{-}$ is 0.20 eV in LiNH$_{2}$ or 0.29 eV in LiBH$_{4}$.\cite{hoang2011amideAC,hoang2011amidePRB,hoang2011LiBH4}

Figure~\ref{fig2}(c) shows the structure of (Li$_{i}^{+}$,$V_{\mathrm{Li}}^{-}$) Frenkel pair. The distance between Li$_{i}^{+}$ and $V_{\mathrm{Li}}^{-}$ is 2.96 {\AA}. (Li$_{i}^{+}$,$V_{\mathrm{Li}}^{-}$) has a formation energy of 0.55 eV and a binding energy of 0.58 eV. For comparison, the calculated formation energies of a similar Frenkel pair in LiNH$_{2}$ and LiBH$_{4}$ are 0.65 eV and 0.95 eV, respectively.\cite{hoang2011amidePRB,hoang2011amidePRB,hoang2011LiBH4}

\subsection{Boron-related defects}

\begin{figure}
\begin{center}
\includegraphics[width=3.0in]{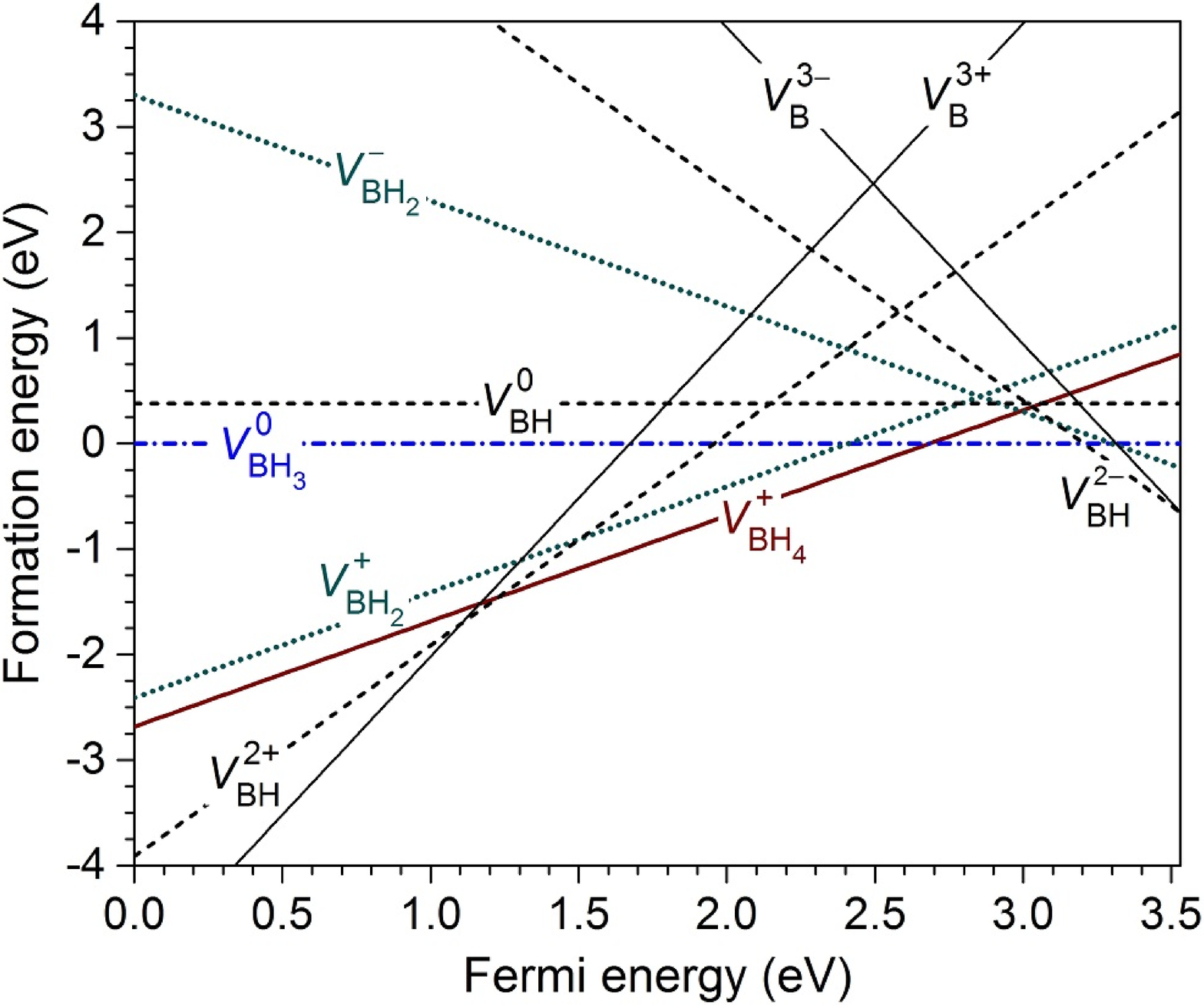}
\end{center}
\vspace{-0.2in}
\caption{Calculated formation energies of boron-related defects, plotted as a function of Fermi energy with respect to the VBM.}\label{LiBNH;FE;B} 
\end{figure}

Figure \ref{LiBNH;FE;B} shows the calculated formation energies of $V_{\mathrm{B}}$, $V_{\mathrm{BH}}$, $V_{\mathrm{BH_{2}}}$, $V_{\mathrm{BH_{3}}}$, and $V_{\mathrm{BH_{4}}}$ in different charge states. Like in LiBH$_{4}$,\cite{hoang2011LiBH4} the creation of $V_{\mathrm{BH_{4}}}^{+}$ involves removing an entire (BH$_{4}$)$^{-}$ unit from the bulk. We find that there is very small change in the local lattice structure surrounding this defect. The structure and energetics of other boron-related defects can be interpreted in terms of $V_{\mathrm{BH_{4}}}^{+}$ and hydrogen-related defects such as H$_{i}^{+}$, H$_{i}^{-}$, and/or (H$_{2}$)$_{i}$, similar to our analysis of the defects in LiBH$_{4}$.\cite{hoang2011LiBH4} For example, $V_{\mathrm{BH_{3}}}^{0}$ can be regarded as a complex of $V_{\mathrm{BH_{4}}}^{+}$ and H$_{i}^{-}$ with a binding energy of 0.91 eV. Finally, the boron interstitial (B$_{i}$; not included in Fig.~\ref{LiBNH;FE;B}) is most stable as B$_{i}^-$ whose formation energy is 2.19 eV at $\mu_{e}=2.92$ eV. The structure of B$_{i}^-$ consists of a BH$_{2}$ unit and two NH units, i.e., an HN$-$BH$_{2}$$-$NH complex, with the B$-$N distance being 1.59 {\AA} (compared to that of 1.35 {\AA} in Li$_{3}$BN$_{2}$). Other charge states of B$_{i}$ have much higher formation energies.

The migration of $V_{\mathrm{BH_{4}}}^{+}$ involves moving a nearby (BH$_{4}$)$^{-}$ unit to the vacancy, with an energy barrier of 0.19 eV. For $V_{\mathrm{BH_{3}}}^{0}$, which can be considered as a complex of $V_{\mathrm{BH_{4}}}^{+}$ and H$_{i}^{-}$, the lower bound of the migration barrier is 0.49 eV. For comparison, the migration barrier of $V_{\mathrm{BH_{4}}}^{+}$ in LiBH$_{4}$ is 0.27 eV.\cite{hoang2011LiBH4}

\subsection{Nitrogen-related defects}

\begin{figure}
\begin{center}
\includegraphics[width=3.0in]{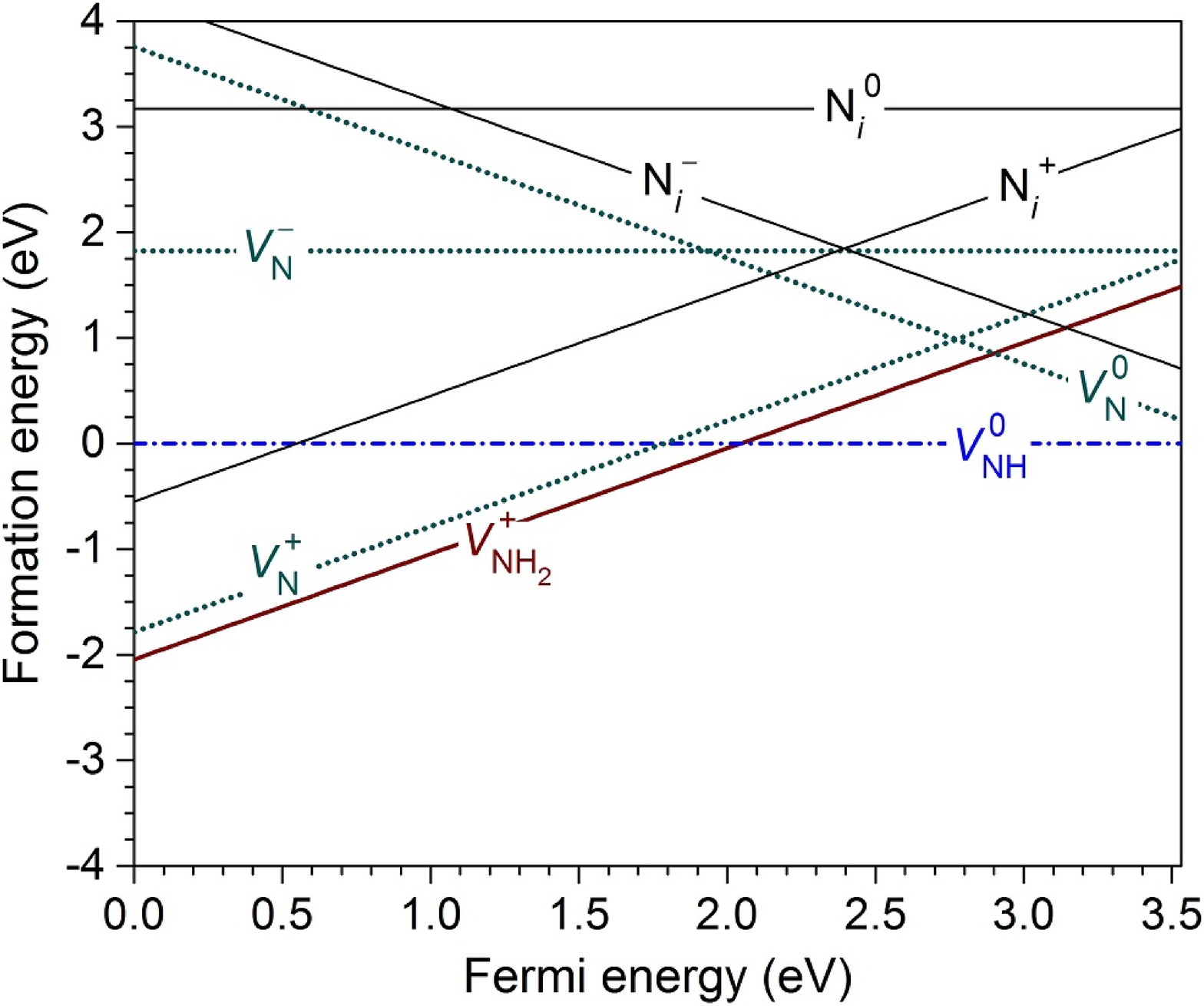}
\end{center}
\vspace{-0.2in}
\caption{Calculated formation energies of nitrogen-related defects, plotted as a function of Fermi energy with respect to the VBM.}\label{fig5} 
\end{figure}

Figure \ref{fig5} shows the calculated formation energies of NH$_{2}$ vacancies ($V_{\mathrm{NH_{2}}}$), NH vacancies ($V_{\mathrm{NH}}$), and nitrogen vacancies ($V_{\mathrm{N}}$) and interstitials (N$_{i}$). We find that $V_{\mathrm{NH_{2}}}$ is energetically stable as $V_{\mathrm{NH_{2}}}^{+}$, $V_{\mathrm{NH}}$ as $V_{\mathrm{NH}}^0$, $V_{\mathrm{N}}$ as $V_{\mathrm{N}}^+$ and $V_{\mathrm{N}}^-$, and N$_{i}$ as N$_{i}^{+}$ and N$_{i}^{-}$ configurations. The creation of $V_{\mathrm{NH_{2}}}^{+}$ corresponds to the removal of an entire (NH$_{2}$)$^{-}$ unit from the bulk. We find that there is very small change in the local lattice structure surrounding this defect. The formation of $V_{\mathrm{NH}}^{0}$, on the other hand, leaves one H atom in the void left by a removed (NH$_{2}$)$^{-}$ unit. $V_{\mathrm{NH}}^{0}$ can be then regarded as a complex of $V_{\mathrm{NH_{2}}}^{+}$ and H$_{i}^{-}$, with a binding energy of 1.53 eV. Similarly, $V_{\mathrm{N}}^{+}$ can be regarded as a complex composed of $V_{\mathrm{NH_{2}}}^{+}$ and (H$_{2}$)$_{i}$, and $V_{\mathrm{N}}^{-}$ as a complex of $V_{\mathrm{NH_{2}}}^{+}$ and two H$_{i}^{-}$ defects. The structure and energetics of these nitrogen-related defects are, therefore, similar to those in LiNH$_{2}$.\cite{hoang2011amidePRB} For nitrogen interstitials, the structure of N$_{i}^{0}$ and N$_{i}^{-}$ consists of a BH$_{3}$ unit and an NH unit, i.e., an H$_{3}$B$-$NH complex, with the B$-$N distance being 1.48 {\AA} in N$_{i}^{0}$ or 1.50 {\AA} in N$_{i}^{-}$. The structure of N$_{i}^{+}$, on the other hand, consists of a BH$_{2}$ unit and an NH$_{2}$ unit, i.e., an H$_{2}$B$-$NH$_{2}$ complex, with the B$-$N distance being 1.39 {\AA}.

The migration of $V_{\mathrm{NH_{2}}}^{+}$ involves moving an NH$_{2}^{-}$ unit to the vacancy, with a calculated energy barrier of 0.59 eV. For $V_{\mathrm{NH}}^{0}$, which can be considered as a complex of $V_{\mathrm{NH_{2}}}^{+}$ and H$_{i}^{-}$, the lower bound of the migration barrier is 0.59 eV. For comparison, the migration barrier of $V_{\mathrm{NH_{2}}}^{+}$ in LiNH$_{2}$ is 0.87 eV.\cite{hoang2011amidePRB}

\section{Discussion}

\begin{table}
\caption{Formation energies ($E^{f}$), migration barriers ($E_{m}$), and binding energies ($E_{b}$) of selected native defects in Li$_{4}$BN$_{3}$H$_{10}$. Migration barriers denoted by an asterisk ($^{\ast}$) are estimated by considering the defect as a complex and taking the highest of the migration barriers of the constituents.}\label{tab:libnh}
\begin{tabular}{cclcc}
\hline
Defect&$E^f$ (eV)& $E_m$ (eV)& $E_{b}$ (eV)&Constituents \\
\hline
H$_{i}^+$ &1.36&0.48&	\\
H$_{i}^-$ &0.58&0.49&	\\
$V_{\rm{H}}^+$ &2.08&0.64& \\
$V_{\rm{H}}^-$ &0.98&1.02&	\\
(H$_{2}$)$_{i}$	&0.75&&	\\
Li$_{i}^+$		 &0.73&0.43&	\\
$V_{\rm{Li}}^-$	&0.40&0.20&	\\
Li$_{\rm{H}}^0$	&0.46&1.02$^{\ast}$&1.25&Li$_{i}^{+}$ + $V_{\rm{H}}^{-}$	\\
$V_{\rm{NH}_{2}}^{+}$ &0.94&0.59&	\\
$V_{\rm{NH}}^0$	 &0.00&0.59$^{\ast}$&1.53&$V_{\rm{NH_{2}}}^{+}$ + H$_{i}^{-}$	\\
$V_{\rm{BH_4}}^+$	&0.32&0.19&	\\
$V_{\rm{BH}}^0$	&0.38&&1.27&$V_{\rm{BH}_{4}}^{+}$ + (H$_{2}$)$_{i}$ + H$_{i}^{-}$	\\
$V_{\rm{BH_2}}^-$	&0.32&0.49$^{\ast}$&1.17&$V_{\rm{BH}_{4}}^{+}$ + 2H$_{i}^{-}$	\\
$V_{\rm{BH_{3}}}^0$	&0.00&0.49$^{\ast}$&0.91&$V_{\rm{BH}_{4}}^{+}$ + H$_{i}^{-}$	\\
\hline
\end{tabular}
\end{table}

We list in Table~\ref{tab:libnh} formation energies and migration barriers of native defects and defect complexes that are most relevant to  lithium-ion conduction and decomposition of Li$_{4}$BN$_{3}$H$_{10}$. The formation energies for charged defects are taken at $\mu_{e}=2.92$ eV (hereafter referred to as $\mu_{e}^{\rm int}$), where the charge neutrality condition is maintained. This Fermi-energy position is determined by solving self-consistently the charge neutrality equation that involves the concentrations of all native (intrinsic) defects,\cite{walle:3851} assuming that electrically active impurities are absent or occur with much lower concentrations than the charged native defects. In this case, it is determined exclusively by $V_{\rm{BH_4}}^+$ and $V_{\rm{BH_2}}^-$ (i.e., a complex of $V_{\rm{BH}_{4}}^{+}$ and two H$_{i}^{-}$), two defects that have the lowest formation energies. For comparison, Farrell and Wolverton reported $\mu_{e}^{\rm int}\sim$2.5$-$3.2 eV under different sets of the atomic chemical potentials.\cite{farrell2012}

It emerges from our results that some native defects in Li$_{4}$BN$_{3}$H$_{10}$ can have very low formation energies. With our choice of the atomic chemical potentials, $V_{\rm{NH}}^0$ (a complex of $V_{\rm{NH_{2}}}^{+}$ and H$_{i}^{-}$) and $V_{\rm{BH_{3}}}^0$ (a complex of $V_{\rm{BH}_{4}}^{+}$ and H$_{i}^{-}$) even have a zero formation energy. Farrell and Wolverton also found very low formation energies for these defects.\cite{farrell2012} The elementary defects $V_{\rm{NH_{2}}}^{+}$, $V_{\rm{BH}_{4}}^{+}$, and H$_{i}^{-}$ that make up the neutral complexes also have low formation energies; {\it cf.} Table~\ref{tab:libnh}. Overall, our results are in qualitative agreement with those reported by Farrell and Wolverton,\cite{farrell2012} and consistent with our results for native defects in LiNH$_{2}$ and LiBH$_{4}$ reported previously.\cite{hoang2011amideAC,hoang2011amidePRB,hoang2011LiBH4}

\subsection{Lithium-ion conduction}

The calculated formation energy of the (Li$_{i}^{+}$,$V_{\mathrm{Li}}^{-}$) Frenkel pair is only 0.55 eV, much lower than that of the hydrogen Frenkel pairs. The low formation energy of (Li$_{i}^{+}$,$V_{\mathrm{Li}}^{-}$) suggests that Li$_{4}$BN$_{3}$H$_{10}$ is prone to Frenkel disorder on the Li sublattice. Farrell {\it et al.},\cite{Farrell2009} through first-principles molecular dynamics simulations, also found that the Li sublattice disorders before the anionic sublattices and the energy barrier of Li migration is $\sim$0.21 eV in a temperature region above the experimental melting point. Experimentally, Matsuo {\it et al.}\cite{Matsuo2009} reported an activation energy of 0.26 eV for the ionic conduction in Li$_{4}$BN$_{3}$H$_{10}$ before melting. These values are very close to our calculated value (0.20 eV) for the migration barrier of $V_{\rm{Li}}^-$. In general, the activation energy for ionic conduction is the sum of the formation energy and migration barrier, i.e., $E_{a} = E^{f} + E_{m}$. However, it is very likely that in the measurements of the ionic conductivity in Ref.~\cite{Matsuo2009} the Li sublattice was already disordered and there were plenty of {\it athermal} Li vacancies and interstitials. In that case, the activation energy is dominated by the migration barrier term, i.e., $E_{a} \sim E_{m}$,\cite{Hoang2014} which explains why our calculated migration barrier of $V_{\mathrm{Li}}^{-}$ is comparable to the measured activation energy.

\subsection{Decomposition mechanism}

Let us now discuss the role of native defects in the decomposition of Li$_{4}$BN$_{3}$H$_{10}$. Like in LiNH$_{2}$ and LiBH$_{4}$,\cite{hoang2011amideAC,hoang2011amidePRB,hoang2011LiBH4} it is important to note that the decomposition involves breaking N$-$H bonds in the (NH$_{2}$)$^{-}$ units and B$-$H bonds in the (BH$_{4}$)$^{-}$ units, which can be accomplished through the creation of relevant native defects. Besides, the process necessarily involves hydrogen, boron, and/or nitrogen mass transport in the bulk mediated by native defects; and as charged defects are migrating, local and global charge neutrality must be maintained. Finally, charge and mass conservation is required for the creation of defects in the interior of the material.

$V_{\rm{H}}^-$, for instance, can form in the interior of the material via the (H$_{i}^{+}$,$V_{\mathrm{H}}^{-}$) Frenkel pair mechanism [Fig.~\ref{fig2}(a)] in which both charge and mass are conserved, or at the surface or interface. $V_{\rm{H}}^+$, on the other hand, is not likely to form in the bulk because the formation energy of the (H$_{i}^{-}$,$V_{\mathrm{H}}^{+}$) Frenkel pair [Fig.~\ref{fig2}(b)] is relatively high (2.14 eV), but it certainly can form at the surface or interface. $V_{\rm{NH}_{2}}^{+}$ and $V_{\rm{BH_4}}^+$ can only be created at the surface or interface since the creation of such defects inside the material requires creation of corresponding (NH$_{2}$)$^{-}$ and (BH$_{4}$)$^{-}$ interstitials which are too high in energy. Finally, Li$_{i}^+$ and $V_{\rm{Li}}^-$ can easily form in the bulk through the lithium Frenkel pair mechanism [Fig.~\ref{fig2}(c)]. With their low formation energies and high mobilities, these lithium interstitial and vacancy can act as accompanying defects in mass transport, providing local charge neutrality as hydrogen-, boron-, and nitrogen-related charged defects migrating in the bulk.

Given the above considerations and the properties of the defects, Li$_{4}$BN$_{3}$H$_{10}$ decomposition can be described in terms of the following processes which may occur simultaneously:

(i) $V_{\rm{H}}^{-}$ is created at the surface or interface by removing an H$^{+}$ from the bulk. This H$^{+}$ ion can combine with H$^{-}$ [that is liberated from Li$_{4}$BN$_{3}$H$_{10}$ when creating $V_{\rm{BH_4}}^+$ via process (ii), see below] to form H$_{2}$, or with a surface (NH$_{2}$)$^{-}$ unit to form NH$_{3}$ that is subsequently released or reacts with other species (see below). In order to maintain the reaction, H$^{+}$ has to be transported to the surface/interface, which is equivalent to $V_{\rm{H}}^{-}$ diffusing into the bulk. As $V_{\rm{H}}^{-}$ is migrating, local charge neutrality is maintained by the mobile Li$_{i}^+$. These two defects can combine and form Li$_{\rm{H}}^0$, which is in fact a Li$_{2}$NH unit inside Li$_{4}$BN$_{3}$H$_{10}$. We note that $V_{\rm{H}}^{-}$ can also be created simultaneously with H$_{i}^{+}$ in the interior of the material through forming a (H$_{i}^{+}$,$V_{\mathrm{H}}^{-}$) Frenkel pair. $V_{\rm{H}}^{-}$ and H$_{i}^{+}$ then become separated as H$_{i}^{+}$ jumps from one (NH$_{2}$)$^{-}$ unit to another. This is equivalent to displacing the NH$_{3}$ unit away from the (NH)$^{2-}$ unit, leaving two Li$^{+}$ next to (NH)$^{2-}$, i.e., a formula unit of Li$_{2}$NH. H$_{i}^{+}$ then migrates to the surface/interface and is released as NH$_{3}$. These are the same mechanisms we have proposed for the decomposition of LiNH$_{2}$ into Li$_{2}$NH and NH$_{3}$ as described in Refs.~\cite{hoang2011amideAC} and \cite{hoang2011amidePRB}. The mechanism happening at the surface/interface is expected to be dominant over the bulk mechanism since the energy required for the creation of defects inside the material is higher.

Alternatively, one starts with the creation of $V_{\rm{NH}_{2}}^{+}$ at the surface or interface by removing one (NH$_{2}$)$^{-}$ unit from the bulk. This unit then can combines with one hydrogen atom from a surface (NH$_{2}$)$^{-}$ unit and releases as NH$_{3}$. This process also leaves Li$_{4}$BN$_{3}$H$_{10}$ with a $V_{\rm{H}}^-$ near the surface/interface. In order to maintain the reaction, (NH$_{2}$)$^{-}$ has to be transported to the surface/interface, which is equivalent to $V_{\rm{NH}_{2}}^{+}$ diffusing into the bulk. As $V_{\rm{NH}_{2}}^{+}$ is migrating, local charge neutrality is maintained by having the highly mobile $V_{\rm{Li}}^-$ in the vacancy's vicinity. The newly created $V_{\rm{H}}^-$ also needs to diffuse into the bulk. As this vacancy is migrating, local charge neutrality is maintained by the mobile Li$_{i}^+$. These two defects can combine and form Li$_{\rm{H}}^0$, which is in fact a Li$_{2}$NH unit. This description is thus equivalent to the above mechanism that starts with the creation of $V_{\rm{H}}^{-}$ at the surface/interface. In both descriptions, the creation of $V_{\rm{H}}^-$ is crucial since it is responsible for breaking N$-$H bonds and turning (NH$_{2}$)$^{-}$ into (NH)$^{2-}$.

(ii) $V_{\rm{BH_4}}^+$ is created at the surface or interface by removing a (BH$_{4}$)$^{-}$ unit from the bulk. Since (BH$_{4}$)$^{-}$ is not stable outside the material, it dissociates into BH$_{3}$ and H$^{-}$ where the latter stays near the surface/interface. This process is similar to that for LiBH$_{4}$ decomposition as described in Ref.~\cite{hoang2011LiBH4}. The BH$_{3}$ unit can then combine with the NH$_{3}$ unit [that is released from the bulk through process (i)] to form ammonia borane (H$_{3}$NBH$_{3}$) or some other intermediates which subsequently release H$_{2}$ and act as nucleation sites for the formation of Li$_{3}$BN$_{2}$. We note that the amount of NH$_{3}$ can be three times higher than that of BH$_{3}$ because the number of (NH$_{2}$)$^{-}$ units is three times higher than that of (BH$_{4}$)$^{-}$ units in Li$_{4}$BN$_{3}$H$_{10}$. From the surface/interface, H$^{-}$ combines with H$^{+}$ [that is liberated from Li$_{4}$BN$_{3}$H$_{10}$ when creating $V_{\rm{H}}^{-}$ via process (i)] to form H$_{2}$, or diffuses into the bulk in form of H$_{i}^-$. In the latter case, H$_{i}^-$ can then combine with Li$^{+}$ to form LiH, which can be a product or an intermediate for further reactions. The hydrogen interstitial can also diffuses along with $V_{\rm{BH_4}}^+$ in form of $V_{\rm{BH}_3}^0$. Like in process (i) that is associated with the formation and migration of $V_{\rm{H}}^{-}$, the mobile Li$_{i}^+$ and $V_{\rm{Li}}^-$ provide local charge neutrality as $V_{\rm{BH_4}}^+$ and/or H$_{i}^-$ are migrating in the bulk.

The rate-limiting step in (i) and (ii) is not the creation of defects at the surface or interface, but the diffusion of $V_{\rm{H}}^-$, H$_{i}^-$, $V_{\rm{NH}_{2}}^{+}$, $V_{\rm{BH_4}}^+$, or (Li$_{i}^+$,$V_{\rm{Li}}^-$) inside the material, whichever defect that has the highest activation energy for formation and migration. The only assumption here is that the formation energy of these defects on the surface/interface is lower than in the bulk, which is a safe assumption given the bonding environment at the surface/interface is less constrained than in the bulk. Since these defects, except for the lithium Frenkel pair, are charged, their formation energies and hence concentrations are dependent on the Fermi-energy position. This opens the door to manipulating their concentrations and hence the decomposition kinetics through shifting the position of the Fermi energy, which can be accomplished by, e.g., incorporating suitable electrically active impurities into the system.\cite{hoang_prb_2009,hoang2011LiBH4}

\subsection{Effects of metal additives}

As reported previously,\cite{hoang_prb_2009} some transition-metal impurities such as Ni, Pd, and Pt can be electrically active in Li$_{4}$BN$_{3}$H$_{10}$ and effective in shifting the Fermi energy. When incorporated into Li$_{4}$BN$_{3}$H$_{10}$ at a certain lattice site with a concentration higher than that of the charged native defects, often through non-equilibrium processes such as high-energy ball milling as noted in Ref.~\cite{hoang2011LiBH4}, these impurities determine the Fermi energy of the system and shift it to a new position (hereafter referred to as $\mu_{e}^{\rm ext}$, the Fermi-energy position determined by the extrinsic defects). Specifically, Ni can shift the Fermi energy to $\mu_{e}^{\rm ext}$ at 1.91 eV (if incorporated on the B site), 2.52 eV (N site ), or 1.87 eV (Li site); 1.82 eV (B site), 2.15 eV (N site), or 1.73 eV (Li site) for Pd; and 1.78 eV (B site), 2.21 eV (N site), or 1.84 eV (Li site) for Pt. Ni, Pd, and Pt are not effective in shifting the Fermi energy if incorporated at interstitial sites.\cite{hoang_prb_2009} For all these impurities, $\mu_{e}^{\rm ext}$ is much lower than $\mu_{e}^{\rm int}$, i.e., the Fermi energy is shifted toward the VBM, thus lowering (increasing) the formation energy of positively (negatively) charged native defects. The incorporation of Ni, Pd, or Pt thus increases the activation energy associated with $V_{\rm{H}}^-$, i.e., delaying the formation of NH$_{3}$, and decreases the activation energy associated with $V_{\rm{BH_4}}^+$, i.e., enhancing the formation of BH$_{3}$ and hence H$_{2}$ and/or intermediates for H$_{2}$ release and lowering the dehydrogenation temperature. Delaying the formation and subsequent release of NH$_{3}$ has important consequences: it enhances the probability of NH$_{3}$ [created in process (i)] being captured by BH$_{3}$ [created in process (ii)] or other species before being released as NH$_{3}$ gas. Our results thus explain why metal additives such as NiCl$_{2}$, Pd (or PdCl$_{2}$), and Pt (or PtCl$_{2}$) are effective in suppressing the release of NH$_{3}$ gas from the decomposition of Li$_{4}$BN$_{3}$H$_{10}$ and lowering the dehydrogenation temperature.

\section{Conclusions}

We have carried out a comprehensive first-principles study of native point defects and defect complexes in Li$_{4}$BN$_{3}$H$_{10}$. We find that lithium interstitials and vacancies are highly mobile and can be created in the interior of the material via a Frenkel pair mechanism with a low formation energy. These defects can participate in lithium-ion conduction or act as accompanying defects which provide local charge neutrality in hydrogen, boron, or nitrogen mass transport. We have proposed an atomistic mechanism for the decomposition of Li$_{4}$BN$_{3}$H$_{10}$, involving the formation and migration of $V_{\rm{H}}^-$, H$_{i}^-$, $V_{\rm{NH}_{2}}^{+}$, $V_{\rm{BH_4}}^+$, and (Li$_{i}^+$,$V_{\rm{Li}}^-$) in the bulk. On the basis of this mechanism, we explain the decomposition and dehydrogenation of Li$_{4}$BN$_{3}$H$_{10}$ and the effects of metal additives on these processes and, particularly, the suppression of NH$_{3}$ release and the lowering of the dehydrogenation temperature as observed in experiments.

\begin{acknowledgments}

This work was supported by the Office of Science of the U.S.~Department of Energy (Grant No.~DE-FG02-07ER46434) and by the Center for Computationally Assisted Science and Technology (CCAST) at North Dakota State University. High performance computing resources were provided by CCAST, the Texas Advanced Computing Center (TACC) at the University of Texas at Austin, and the National Energy Research Scientific Computing Center (NERSC), a DOE Office of Science User Facility supported by the Office of Science of the U.S. Department of Energy under Contract No.~DE-AC02-05CH11231.

\end{acknowledgments}


%

\end{document}